\pdfoutput=1
\documentclass[PAPER, atlasdraft=false, UKenglish, cernpreprint, texmf, orcidlogo]{atlasdoc}
\usepackage{atlaspackage}
\usepackage{atlasbiblatex}

\usepackage{multirow}

\usepackage{atlasphysics}
\usepackage{bm}
\graphicspath{{logos/}{figures/}}

\usepackage{ANA-HMBS-2024-39-PAPER-defs}

\AtlasTitle{Search for Higgs boson decays into a pair of pseudoscalar particles in the $\gamma\gamma\tau_{\text{had}}\tau_{\text{had}}$ final state using $pp$ collisions at $\sqrt{s}=13$ TeV with the ATLAS detector}

\AtlasAbstract{
A search for exotic decays of the 125 GeV Higgs boson into a pair of new spin-0 particles, $H \to aa$, where one decays into a photon pair and the other into a $\tau$-lepton pair, is presented. Hadronic decays of the $\tau$-leptons are considered and reconstructed using a dedicated tagger for collimated $\tau$-lepton pairs. The search uses 140~fb$^{-1}$ of proton--proton collision data at a centre-of-mass energy of $\sqrt{s}=13$ TeV recorded between 2015 and 2018 by the ATLAS experiment at the Large Hadron Collider. The search is performed in the mass range of the $a$ boson between 10 GeV and 60 GeV. No significant excess of events is observed above the Standard Model background expectation. Model-independent upper limits at 95\% confidence level are set on the branching ratio of the Higgs boson to the $\gamma\gamma\tau\tau$ final state, $\mathcal{B}(H\to aa\to \gamma\gamma\tau\tau)$, ranging from 0.2\% to 2\%, depending on the $a$-boson mass hypothesis.}

\AtlasRefCode{HMBS-2024-39}

\PreprintIdNumber{CERN-EP-2024-323}

\arXivId{2412.14046}

\HepDataRecord{157781}

\AtlasJournal{JHEP}
\AtlasJournalRef{JHEP 03 (2025) 190}
\AtlasDOI{10.1007/JHEP03(2025)190}

\AtlasCoverEgroupAnalysisTeam{atlas-hmbs-2024-39-analysis-team@cern.ch}

\hypersetup{pdftitle={ATLAS document},pdfauthor={The ATLAS Collaboration}}

\begin{document}

\maketitle

\tableofcontents

%
%

%

\section{Introduction}

The observation of a Standard Model (SM) like Higgs boson with a mass of 125~\GeV~\cite{HIGG-2012-27,CMS-HIG-12-028} marked the beginning of a new era in physics at the Large Hadron Collider (LHC).
Currently, measurements of the properties of the Higgs boson have shown no significant deviations from the SM predictions, and the existing constraints on the couplings of the Higgs boson to SM particles limit the branching ratio into non-SM or \textit{exotic} decays of the Higgs boson to less than approximately 10\%~\cite{HIGG-2021-23,CMS-HIG-22-001}.
However, several beyond-the-SM scenarios predict a SM-like Higgs boson with a small intrinsic decay width ($\Gamma_H \sim 4$~\MeV) and non-SM decays with a branching ratio of $\mathord{\sim}10\%$
without modifying the couplings of the Higgs boson to SM particles beyond the existing bounds~\cite{Curtin:2013fra}.

Light (pseudo)scalars, referred to here as $a$ bosons, that couple to the 125~\GeV Higgs boson ($H$) can enable new decay modes, and appear in many well-motivated extensions of the SM\@. Examples include theories with an extended Higgs sector~\cite{Dobrescu:2000yn,Ellwanger:2003jt,Derm_ek_2005,Chang:2008cw},
dark matter models~\cite{Silveira:1985rk,Pospelov:2007mp,Ipek:2014gua},
models with a first-order electroweak phase transition~\cite{Profumo:2007wc,Blinov:2015sna},
and theories with neutral naturalness~\cite{Craig:2015pha,Curtin:2015fna}.

This paper presents a search for \signalprocess, with the Higgs boson decaying into two $a$ bosons, where one $a$ decays into a photon pair and the other into a $\tau_{\text{had}}$ pair, considering only the hadronic decays of the $\tau$-lepton. The dataset consists of the proton--proton ($pp$) collisions recorded at $\sqrt{s} = 13$~\TeV~with the ATLAS detector during the LHC Run~2 (2015--2018), corresponding to an integrated luminosity of \integLumi.
Searches targeting final states with either a photon pair or a $\tau$-lepton pair, as decay products of such a (pseudo)scalar, have previously been conducted by both the ATLAS~\cite{HDBS-2021-07,HDBS-2019-19,HDBS-2019-09} and CMS~\cite{CMS-HIG-22-007,CMS-PAS-SUS-24-002,CMS-HIG-21-003,CMS-HIG-22-003b} collaborations.
However, this search presents the first result in the combined \finalstate~final state. The two photons are used to provide a highly efficient trigger selection and an excellent mass resolution, while the two $\tau$-leptons benefit from a large branching ratio in models where the coupling is proportional to the particle mass, such as two-Higgs-doublet models~\cite{Branco:2011iw,Lee:1973iz}.

This analysis uses the diphoton invariant mass, \mgg, as the main observable, further exploring the range below 60~\GeV --- reaching 10 GeV as the lowest mass probed, limited by the standard trigger and photon reconstruction techniques --- studied for the first time in Ref.~\cite{HIGG-2019-23} with $pp$ collisions. A similar strategy is implemented, requiring a boosted photon pair with transverse momentum \ptgg\ larger than 50~\GeV,
which eases the description of the background shape with analytical functions.
The requirement of an additional pair of hadronically decaying $\tau$-leptons further increases the background rejection and new reconstruction techniques for boosted $\tau$-lepton pairs are implemented to improve the sensitivity in the low-mass regime.

The paper is structured as follows. A brief discussion of the ATLAS detector and an overview of the Monte Carlo samples and data used in the analysis are presented in Sections~\ref{sec:detector} and~\ref{sec:data_mc}. The event reconstruction and selection are described in Section~\ref{sec:selection}, with special focus on the custom reconstruction of boosted \ditau\ signals. The signal modelling and background estimates are discussed in Section~\ref{sec:modelling}, followed by a description of the dominant systematic uncertainties and the statistical method in Section~\ref{sec:stats}. Finally, the results are presented in Section~\ref{sec:results} and the conclusions are drawn in Section~\ref{sec:conclusion}.

%

 %

%
\section{ATLAS detector}
\label{sec:detector}


%
\newcommand{\AtlasCoordFootnote}{%
ATLAS uses a right-handed coordinate system with its origin at the nominal interaction point (IP) in the centre of the detector and the \(z\)-axis along the beam pipe. The \(x\)-axis points from the IP to the centre of the LHC ring, and the \(y\)-axis points upwards. Polar coordinates \((r,\phi)\) are used in the transverse plane, \(\phi\) being the azimuthal angle around the \(z\)-axis. The pseudorapidity is defined in terms of the polar angle \(\theta\) as \(\eta = -\ln \tan(\theta/2)\) and is equal to the rapidity $y=\frac{1}{2}\ln\left(\frac{E+p_{z}}{E-p_{z}}\right)$ in the relativistic limit. Angular distance is measured in units of \(\Delta R \equiv \sqrt{(\Delta y)^{2} + (\Delta\phi)^{2}}\).}

The ATLAS experiment~\cite{PERF-2007-01} at the LHC is a multipurpose particle detector
with a forward--backward symmetric cylindrical geometry and a near \(4\pi\) coverage in
solid angle.\footnote{\AtlasCoordFootnote}
It consists of an inner tracking detector surrounded by a thin superconducting solenoid
providing a \qty{2}{\tesla} axial magnetic field, electromagnetic and hadronic calorimeters, and a muon spectrometer.
The inner tracking detector covers the pseudorapidity range \(|\eta| < 2.5\).
It consists of silicon pixel, silicon microstrip, and transition radiation tracking detectors.
Lead/liquid-argon (LAr) sampling calorimeters provide electromagnetic (EM) energy measurements
with high granularity within the region \(|\eta|< 3.2\).
A steel/scintillator-tile hadronic calorimeter covers the central pseudorapidity range (\(|\eta| < 1.7\)).
The endcap and forward regions are instrumented with LAr calorimeters
for EM and hadronic energy measurements up to \(|\eta| = 4.9\).
The muon spectrometer surrounds the calorimeters and is based on
three large superconducting air-core toroidal magnets with eight coils each.
The field integral of the toroids ranges between \num{2.0} and \qty{6.0}{\tesla\metre}
across most of the detector.
The muon spectrometer includes a system of precision tracking chambers up to \(|\eta| = 2.7\) and fast detectors for triggering up to \(|\eta| = 2.4\).
The luminosity is measured mainly by the LUCID--2~\cite{LUCID2} detector, which is located close to the beampipe.
A two-level trigger system is used to select events~\cite{TRIG-2016-01}.
The first-level trigger is implemented in hardware and uses a subset of the detector information
to accept events at a rate below \qty{100}{\kHz}.
This is followed by a software-based trigger that
reduces the accepted event rate to \qty{1}{\kHz} on average
depending on the data-taking conditions.
A software suite~\cite{SOFT-2022-02} is used in data simulation, in the reconstruction
and analysis of real and simulated data, in detector operations, and in the trigger and data acquisition
systems of the experiment.


%
\section{Data and simulated event samples}
\label{sec:data_mc}

The search is performed using the $13$~\TeV~$pp$ collision dataset collected from 2015 to 2018 by the ATLAS detector, referred to as the full Run~2 dataset in the following.
Only events with stable beam conditions and all ATLAS subsystems operational are considered~\cite{DAPR-2018-01},
corresponding to an integrated luminosity of 140~\ifb with an uncertainty of 0.83\%~\cite{DAPR-2021-01}.
The data were recorded using a set of diphoton triggers~\cite{TRIG-2018-05} that required two electromagnetic energy clusters satisfying identification criteria based on their expected shower shapes and transverse energies (\ET) above a certain threshold that varied across the data-taking period to cope with the increase in instantaneous luminosity over the years.
In 2015 and the first portion of 2016, the trigger required a \ET~threshold of 20~\GeV, while in the remainder of 2016 the threshold was set to 22~\GeV.
During 2017 and 2018, the \ET~threshold was set to 20~\GeV\ along with isolation requirements.
An alternative dataset is used for background estimation, collected with a prescaled diphoton trigger with looser identification criteria for the EM shower shapes and a \ET~threshold of 20~\GeV, corresponding to an integrated luminosity of $6.3$~\ifb.

Monte Carlo (MC) simulated samples are used to model signal and background processes, as well as to derive related modelling uncertainties.
The generated signal events are processed through a full simulation~\cite{SOFT-2010-01} of the ATLAS geometry and response based on \GEANT~\cite{Agostinelli:2002hh}, while the background events are processed through a faster simulation where the full simulation of the calorimeter is replaced with a parameterisation of the calorimeter response~\cite{ATL-PHYS-PUB-2010-013}.
The effects of multiple $pp$ interactions in the same bunch crossing as the hard scatter and in neighbouring ones (defined as pile-up) are included using simulated events generated with \PYTHIA[8.186]~\cite{Sjostrand:2007gs} using the A3 set of tuned parameters~\cite{ATL-PHYS-PUB-2016-017} and the \NNPDF[2.3lo]~\cite{Ball:2012cx} parton distribution functions (PDF) set.
Simulated events are weighted to reproduce the distribution of the average number of interactions per bunch crossing observed in data. All simulated events are reconstructed with the same reconstruction algorithms as those used for data.

Signal event samples are generated for the SM Higgs boson produced through gluon--gluon fusion (ggF), using the NNLOPS approach~\cite{Hamilton:2013fea} with \POWHEGBOX~v2~\cite{Nason:2009ai,Alioli:2010xd,Nason:2004rx,Frixione:2007vw} interfaced with \PYTHIA[8.245]~\cite{Sjostrand:2014zea} using the AZNLO set of tuned parameters~\cite{STDM-2012-23} to simulate parton showering, hadronisation and the full decay chain, \signalprocessnohad, where the $\tau$-leptons are set to decay hadronically.
The next-to-next-to-leading-order (NNLO) PDF4LHC15~\cite{Butterworth:2015oua} sets of PDFs were used.
To study on-shell decays of the Higgs boson, the mass of the pseudoscalar resonance, \ma, is varied in 5~\GeV~steps within the range 10~\GeV\ to 60~\GeV. The decay width of the hypothetical resonance is set to 4~\MeV, consistent with the narrow-width approximation, as this value is negligible compared to the experimental resolution.
Only the ggF process --- the dominant Higgs boson production mechanism, accounting for approximately 87\% of the total cross-section --- is considered in this study, ignoring the contributions from the other Higgs boson production modes.

The main background in this search originates from events with two prompt photons and associated jets, which are simulated using the \SHERPA[2.2.4]~\cite{Bothmann:2019yzt} event generator. The matrix elements are calculated at next-to-leading-order (NLO) in QCD for up to one additional parton and at leading-order (LO) in QCD for two or three partons, and are merged with the \SHERPA parton shower simulation using the \MEPSatNLO prescription~\cite{Hoeche:2011fd,Hoeche:2012yf,Catani:2001cc,Hoeche:2009rj}.
The \NNPDF[3.0nnlo]~\cite{Ball:2014uwa} PDF set was used in conjunction with a dedicated parton-shower tune in the \SHERPA generator~\cite{Schumann:2007mg}. Processes with two photons and two $\tau$-leptons in the final state, originating at lowest order from SM $Z$-boson decays, are not considered in the analysis, as the cross-section is negligible compared with the dominant background.

Interference effects between the resonant signal and the background processes are expected to be small for narrow-width signals and are neglected.


%
\section{Event reconstruction and selection}
\label{sec:selection}


%
\subsection{Event reconstruction}

Events are required to contain at least one reconstructed $pp$ collision vertex candidate with at least two associated ID tracks with transverse momenta ($\pt$) larger than $0.5$~\GeV~\cite{ATL-PHYS-PUB-2015-026}.
The primary vertex (PV) for each event is chosen as the reconstructed vertex with the highest sum of the $\pt^2$ of its associated tracks.

Photon candidates are reconstructed from topological clusters of energy deposited in the EM calorimeter, as well as from charged particle vertex and vertices reconstructed in the inner detector originating from photon conversions, and they are calibrated as described in Ref.~\cite{EGAM-2018-01}.
The properties of the EM clusters associated with the two highest-\ET~photons is used to correct the photon direction, resulting in improved diphoton mass resolution.
To reduce the background from jets, photon candidates are required to satisfy \emph{Tight} identification criteria based on the shape of EM showers in the LAr calorimeter and energy leakage into the hadronic calorimeter~\cite{EGAM-2018-01}. The criteria have an identification efficiency that increases with \ET from 70\% at \SI{22}{\GeV} to 90\% above \SI{50}{\GeV}.
Events with one or both photon candidates satisfying a looser identification requirement from the prescaled data are kept for background estimations. To further improve the rejection of jets misidentified as photons, the candidates are required to be isolated using both calorimeter and tracking information.
The calorimeter isolation transverse energy \ETcalo~is required to be smaller than $0.065\times\ET^\gamma$, where \ETcalo~is defined as the sum of the transverse energies of positive-energy topological clusters~\cite{PERF-2014-07} within a cone of size $\Delta R = 0.2$ around the photon candidate, excluding the photon transverse energy $\ET^\gamma$ and correcting for pile-up and underlying-event contributions~\cite{STDM-2010-08,Cacciari:2008gn,Cacciari:2009dp}.
The track isolation transverse energy \ETtrk~is required to be less than $0.05\times\ET^\gamma$, where \ETtrk~is defined as the scalar sum of the transverse momenta of tracks with $\pT > 1$~\GeV\ in a $\Delta R=0.2$ cone around the photon candidate. The tracks considered must satisfy some loose track-quality criteria~\cite{EGAM-2018-01}, should not be associated with the photon conversion vertex if it exists, and originate from the PV set as the diphoton production vertex.
The combined isolation efficiency for pairs of photons fulfilling the identification criteria in simulated signal samples increases with \mgg~from 80\% at \SI{10}{\GeV} to 90\% at \SI{90}{\GeV}.

Electrons and muons are reconstructed to veto events with leptonic final states. Electrons are reconstructed from energy clusters in the EM calorimeter matched to tracks reconstructed in the ID, and are required to have $\pt>27~\GeV$ and $|\eta|<2.47$. Electrons in the calorimeter barrel--endcap transition region ($1.37 <|\eta| < 1.52$) are excluded. Electrons must satisfy the \emph{Tight} identification criteria~\cite{EGAM-2021-02}, the longitudinal impact parameter must be smaller than \SI{0.5}{mm}, and the transverse impact parameter significance smaller than 5, both defined with respect to the beam line. Isolation criteria are applied to the selected electrons using both tracking and calorimeter information with a $\pT$-dependent cone radius~\cite{EGAM-2018-01}.

Muons are reconstructed using different methods depending on the availability of the tracks in the MS and the ID~\cite{MUON-2018-03}, and are required to have $\pT>27~\GeV$ and $|\eta|<2.7$. Similarly to the electrons, muons are required to satisfy the \emph{Medium} identification criteria~\cite{MUON-2018-03}, the longitudinal impact parameter smaller than \SI{0.5}{mm}, and the transverse impact parameter significance smaller than 3. The isolation criteria are based on requirements from both the ID and calorimeters, also with a $\pt$-dependent cone radius~\cite{MUON-2018-03}.

Jets are reconstructed using the \antikt algorithm~\cite{Cacciari:2008gp} as implemented in \textsc{FastJet}~\cite{Fastjet} with a radius parameter $R = 0.4$. The inputs to this algorithm are particle flow objects~\cite{PERF-2015-09}, which combine measurements from the ID and calorimeters~\cite{PERF-2014-07} to improve the jet energy resolution and increase the jet reconstruction efficiency, especially at low jet \pt. The jet energy scale is calibrated to the particle level using simulation and further corrected with in-situ methods~\cite{PERF-2016-04}. Jets are required to have $\pt>25$~\GeV and $|\eta|<2.5$.

The reconstruction of the visible component of hadronically decaying $\tau$-lepton candidates, $\tauhadvis$, is seeded from jets reconstructed using the \antikt algorithm with a radius parameter $R=0.4$ whose inputs are topoclusters, three-dimensional clusters of calorimeter cells~\cite{PERF-2014-07}, calibrated using a local hadronic calibration~\cite{ATL-PHYS-PUB-2015-045}.
Reconstructed $\tauhadvis$ candidates are required to have $\pT>20~\GeV$, $|\eta|<2.5$ (excluding the transition region $1.37 <|\eta| < 1.52$) and either one or three associated tracks with a total charge of $\pm1$~\cite{ATL-PHYS-PUB-2015-045}. The \emph{Loose} identification criteria, applied for the $\tauhadvis$ candidates, are based on a Recurrent Neural Network (RNN) algorithm, which uses tracks and calorimeter clusters associated to $\tauhadvis$ candidates as well as high-level discriminating variables, to reject background from jets misidentified as $\tauhadvis$ candidates~\cite{ATL-PHYS-PUB-2019-033}.
In the following, a \textit{resolved} \ditau\ object refers to two separated $\tauhadvis$ candidates.
The RNN identification is used for resolved $\tauhadvis$ candidates, while a dedicated algorithm for \textit{boosted} \ditau\ objects is described below.

The boosted \ditau\ object refers to two $\tauhadvis$ that are within $\Delta R<0.4$ from each other, such that the reconstruction is performed via a custom algorithm~\cite{TAUP-2020-03} using the substructure of a large-radius jet.
The large-radius jet is reconstructed using the \antikt jet algorithm with a radius parameter $R=1.0$ from particle-flow objects and required to have $\pt>50~\GeV$, followed by the reclustering of its constituents into subjets with $\pT>10~\GeV$, using the \antikt algorithm with $R=0.2$.
At least two subjets are required to define a boosted \ditau\ candidate, with each of the two containing either one or three associated tracks. The subjets are ordered in \pt and referred to as the \textit{leading} and \textit{sub-leading} subjets, respectively.
The charge product of the two leading subjets is $\pm1,$ where the charge of each subjet is defined as the sum of the charges of the associated tracks.
Then a dedicated identification algorithm is implemented to distinguish the boosted \ditau\ candidates from quark- or gluon-initiated jets, using a boosted decision tree (BDT) method, based on a set of kinematic variables. The discriminating variables are derived from track and calorimeter information of the subjets associated with the \ditau\ candidates.
The performance of the reconstruction and identification algorithms is estimated in a calibration analysis, with the measurement of the \ditau\ identification efficiencies in data and simulation, the ratio of which is defined as the boosted \ditau\ \textit{scale factor} (SF). The efficiency measurement is performed via the tag-and-probe technique using boosted \ditau\ objects from the SM $Z\left(\to\tauhad\tauhad\right)+\gamma$ process, recorded with a high-\pt photon trigger.
A \textit{Medium} \ditau\ BDT-based identification criterion is defined, corresponding to a BDT score $>0.35$ selection. The estimated signal efficiency is approximately 70\%, with a measured background rejection factor of about 240. The corresponding SF is $1.00 \pm 0.35\left(\text{stat.}\right) \pm 0.13\left(\text{syst.}\right)$.
The \ditau\ BDT score distributions for both data and simulated signal samples corresponding to \ma\ values of 10~\GeV and 30~\GeV are shown in Figure~\ref{fig:ditau-BDT}.

\begin{figure}
\centering
\includegraphics[width=0.7\columnwidth]{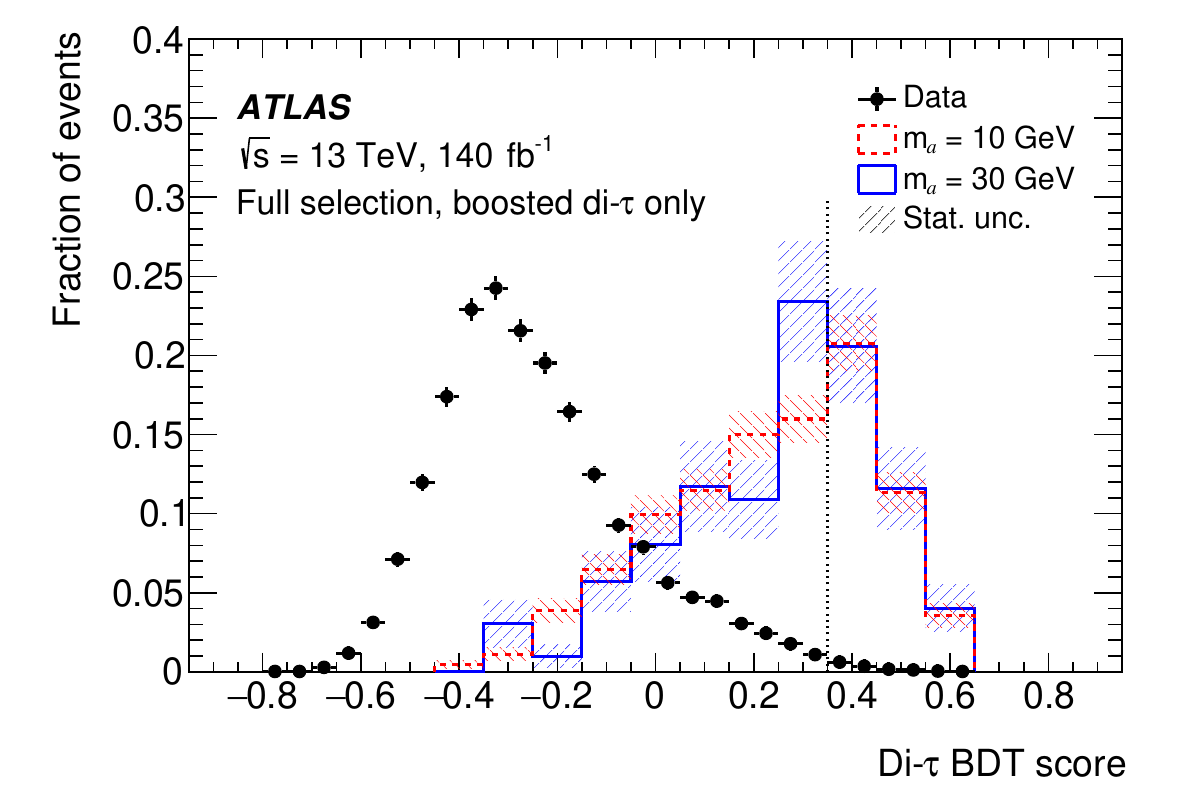}
\caption{
\Ditau\ BDT score distribution for data and two representative signal masspoints, for events that satisfy the selection criteria in Section~\ref{sec:selection} and contain a boosted \ditau\ candidate.
The BDT score $>0.35$ selection (dotted line) is referred to as the \textit{Medium} \ditau\ identification working point and is used in the analysis.
Statistical uncertainties in the data are represented by the errors bars around the points and in the simulated signal by the hatched bands around the histograms.
}\label{fig:ditau-BDT}
\end{figure}

The examination of the visible \pt of the \ditau\ decay products, $p_{\text{T,vis}}^{\tau\tau}$, in the boosted \ditau\ calibration analysis and in this search using simulated events, shows that the former exhibits values an order of magnitude higher.
An extrapolation uncertainty is hence introduced to account for the different kinematic regimes probed in both studies. The boosted \ditau\ SF is extrapolated to the relevant $p_{\text{T,vis}}^{\tau\tau}$ range and is found to be consistent with unity within 50\%, which is taken as a systematic uncertainty and added in quadrature to the boosted \ditau\ SF uncertainty.
\subsection{Event selection}

The event selection requires at least two photon candidates with transverse energies larger than 22~\GeV\ and $|\eta|< 2.37$, excluding the transition regions of the calorimeter, $1.37 <|\eta|< 1.52$. The transverse energy requirement is chosen to mitigate the effects in the trigger efficiency near the trigger thresholds discussed in Section~\ref{sec:data_mc}. The diphoton invariant mass is computed using the transverse energies of the leading and sub-leading photon candidates and their angular separation in both azimuth $\phi$ and pseudorapidity $\eta$, determined from their positions in the calorimeter and the production vertex. An additional kinematic selection is placed on the transverse momentum of the diphoton system, \ptgg, requiring events to have a diphoton pair with $\ptgg>50$~\GeV. This requirement is motivated by the fact that the analysis targets diphoton pairs with low invariant masses, down to about half the trigger energy thresholds, and such pairs are typically highly boosted with respect to the ATLAS detector rest frame. Moreover, the \ptgg~requirement is chosen to reach the best compromise between the statistical uncertainty in the lowest part of the diphoton invariant mass spectrum and sculpting effects on the background shape from the trigger thresholds, whose mismodelling would result in large systematic uncertainties.

The selection additionally requires a boosted \ditau\ candidate or two resolved $\tau$-leptons to provide coverage for the whole \ma\ mass spectrum.
If the event contains a \ditau\ signature that is reconstructed by both the boosted and the resolved \ditau\ algorithms, the resolved $\tau$-lepton pair is selected due to its corresponding \ditau-related identification uncertainties being significantly smaller.
Considering both the resolved and boosted topologies, the \ditau\ tagger recovers sensitivity mostly in the ${\mgg\lesssim25}$~\GeV domain, dominated by the presence of boosted \ditau\ objects, as shown in Figure~\ref{fig:ditau-eff}.
As the \ditau\ signal is expected from a neutral particle, the $\tau$-leptons are required to have opposite charges. As for the identification criteria, the two leading resolved $\tau$-leptons must satisfy the \emph{Loose} identification working point~\cite{ATL-PHYS-PUB-2019-033}, whereas the boosted \ditau\ should fulfil the \emph{Medium} working point~\cite{TAUP-2020-03}. Additionally, the events are required to contain no electrons or muons to reject backgrounds involving leptonic final states, which are not expected from the signal.

\begin{figure}
\centering
\includegraphics[width=0.7\columnwidth]{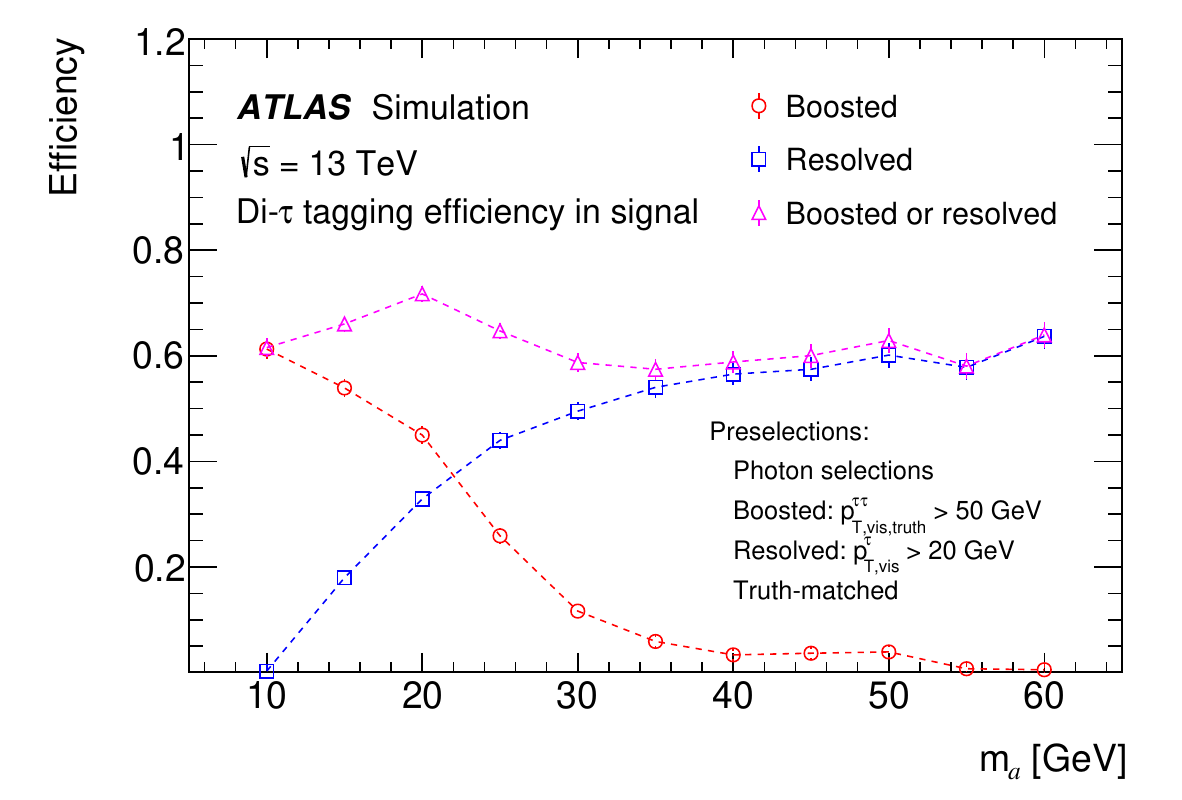}
\caption{
\Ditau\ reconstruction and identification combined efficiency in signal for the different \ditau\ categories: \textit{boosted} (circles), \textit{resolved} (squares) and \textit{boosted or resolved} (triangles). The analysis uses the latter to maximize the signal efficiency across the whole mass range probed. The efficiency is defined using the \textit{truth-matched} preselection that refers to both boosted \ditau\ objects and resolved $\tau$-leptons, geometrically matching the reconstructed object or its constituents to the particle-level object within $\Delta R=0.2$.
Connecting dashed lines are only for visualization purposes.
Only statistical uncertainties are represented.
}\label{fig:ditau-eff}
\end{figure}

Additionally, events that fail the $\tau$-lepton pair requirements, as well as the photon identification and isolation criteria, are kept for background composition studies. For boosted \ditau\ objects, the identification requirement is inverted using a BDT score $< -0.2$ selection, and the two leading subjets are required to have the same charge. In events with a resolved $\tau$-lepton pair, either the two leading $\tau$-leptons are required to have the same charge, or one of them has to fail the \emph{Loose} identification requirement. In the following, these selections are referred to as the \textit{inverted \ditau\ selections}.

%

%
%
%


%
%
%
%

\section{Signal and background modelling}
\label{sec:modelling}


%
\subsection{Signal parameterisation}
\label{sec:signal}
The signal modelling strategy consists of
fitting the diphoton invariant mass distribution of simulated signal samples with different masses to a double-sided Crystal Ball (DSCB) function, composed of a Gaussian core with power-law tails~\cite{HIGG-2014-04,Oreglia:1980cs}.
The parameters of the DSCB function are extracted from these fits for each signal \ma\ hypothesis and parameterised linearly with respect to \ma.
The width of the Gaussian core, driven entirely by the detector resolution, ranges from 0.2 to 1.0~\GeV.
The signal model agrees well with the simulated signal samples, with differences below 2.5\% of the fitted signal yield. This value is used as the signal modelling uncertainty.

\subsection{Background estimation}
\label{sec:bkg}

\subsubsection*{Background decomposition}

The dominant backgrounds generally consist of three components: the continuum diphoton production (\gamgam), photon--jet ($\gamma j$) and jet pair ($jj$) events, where one or more jets are misidentified as photons.
In addition, these background processes contain a $\tau$-lepton pair misidentified from one or more additional jets.
The background consisting of events with genuine photons and hadronically decaying $\tau$-lepton pairs --- originating at lowest order from a $Z$ boson decaying into a $\tau$-lepton pair produced in association with a photon pair --- is highly suppressed due to its small cross-section and the boosted selections, and is estimated to contribute only about 1\% of the data.
This leaves the background to be dominated by the photon processes.

The search makes use of a data-driven background estimate in which the continuum background shape is parameterised by an analytical function. %
Background templates are built to estimate the robustness of the background parameterisation and derive modelling systematic uncertainties.
The nominal background template is composed of the irreducible and reducible sources: the former being the \gamgam~component obtained from simulated diphoton samples; and the latter combining the $\gamma j$ and $jj$ components obtained from control regions, inverting the photon identification criteria, in the alternative data samples. Those are recorded using the prescaled diphoton trigger that allows to relax the photon identification and isolation criteria.
Due to the limited size of the data and simulated samples, both irreducible and reducible sources apply the inverted \ditau\ selections, defined in Section~\ref{sec:selection}. Comparing the diphoton invariant mass distributions obtained for different \ditau\ selections, their shapes are observed to agree within statistical uncertainties, allowing the use of such \ditau\ selections to describe the expected background in data.
The two sources are then combined according to their relative fractions estimated by using the two-dimensional sideband method~\cite{STDM-2012-05}. This method is used in analyses with final states containing two photons, alternately inverting their isolation and identification criteria to measure the contribution of the \gamgam, $\gamma j$ and $jj$ components, as shown in Figure~\ref{fig:bkgDecomp}.
The overall diphoton purity, defined within the \mgg~range 6--68~\GeV as the fraction of the \gamgam~component in the data, is $0.61\pm0.01\text{ (stat.)}\pm0.04\text{ (syst.)}$, in which the systematic uncertainty arises from the different choices of relaxed photon identification criteria.

\begin{figure}
\centering
\includegraphics[width=0.7\textwidth]{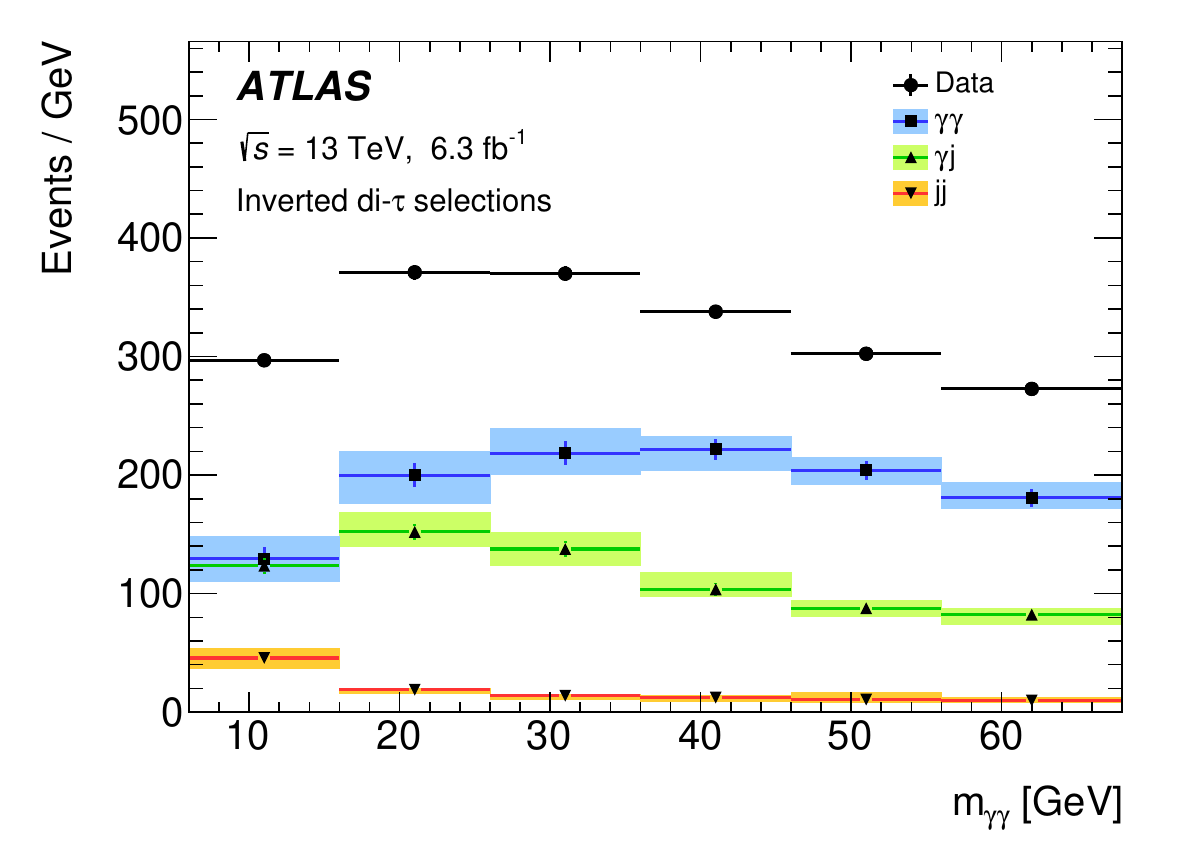}
\hfill
\caption{
Diphoton invariant mass distribution in data after the inverted \ditau\ selections, and its decomposition into contributions from diphoton (\gamgam), photon--jet ($\gamma j$) and jet pair ($jj$) events as determined using the two-dimensional sideband method.
The statistical uncertainties are shown as error bars, while the total uncertainties, including statistical and systematic components added in quadrature, are shown as shaded bands.
}\label{fig:bkgDecomp}
\end{figure}

\subsubsection*{Template shape description}

The entire background template shape is described using an analytic functional form.
The chosen function is a product of a sigmoid function that effectively describes the turn-on at low masses, and an exponential function that models the smoothly falling shape at high masses:
\begin{equation*}
f\left(m_{\gamma\gamma};N, \delta_{\text{sgmd}},\tau_{\text{sgmd}},\lambda_{\text{exp}}\right) = N \times \frac{1}{1+e^{-(m_{\gamma\gamma}-\delta_{\text{sgmd}})/\tau_{\text{sgmd}}}}\times e^{-\lambda_{\text{exp}}\ \mgg},
\end{equation*}
where $N$ is a normalisation constant, $\delta_{\text{sgmd}}$ and $\tau_{\text{sgmd}}$ are the parameters regulating the step of the sigmoid function,
while $\lambda_{\text{exp}}$ is the exponential function decay constant.
The function parameters are fit to the background template in the \mgg\ range 6--68~\GeV.

Variations of the nominal background template are built to validate the flexibility of the function and to estimate systematic uncertainties. The variations are obtained by either modifying the photon identification criteria, varying the \gamgam\ purity by its uncertainty, or altering the \ditau\ identification criteria. The difference in the yield between the various template shape variations and the nominal ranges up to 10\% for different \ditau\ selections and up to 4\% when the photon identification criteria or the normalisation of the \gamgam\ component is varied.

\subsubsection*{Modelling bias evaluation}

The uncertainty arising from the choice of background model is based on signal-plus-background fits to background-only template histograms~\cite{ATL-PHYS-PUB-2020-028}. Any fit signal yield is referred to as a \textit{spurious signal} and is considered as a systematic uncertainty of the background model. The model described has a maximum bias of 13\% of the statistical uncertainty of all the templates.
Although the spurious signal is small, it is susceptible to statistical fluctuations in the data, as the template is built with limited statistics.

To reduce the statistical effects on the background modelling uncertainty, templates are smoothed using a Gaussian Processes regression~\cite{rasmussen}. This is a non-parametric regression technique used to estimate the underlying function of a dataset exploiting the correlation between points separated by a certain distance denominated the \textit{length scale}, without modifying significantly the background shape nor introducing any bias in the yield~\cite{HIGG-2019-23}.
The length scale encodes the minimum feature size expected in the background shape, corresponding in this analysis to the trigger turn-on of size between 1 and 2~\GeV, which is also larger than the signal resolution.
Background templates in the full \mgg\ range 6--68~\GeV are smoothed, removing the statistical fluctuations. Since the smoothing is based on the correlation between neighbouring data, the combined signal and background model fit is performed using the \mgg\ range 10--60~\GeV\ to mitigate any mismodelling introduced by the edge effect. With the statistical fluctuations mitigated, the estimated spurious signal is reduced to less than approximately 5\% of the statistical uncertainty of the template.

The spurious signal is then estimated for all the background template variations described previously. This allows to estimate the bias from the choice of background modelling, which is always largest at the turn-on region and up to 8\% of the expected statistical uncertainty.
The systematic uncertainty results from the envelope of the maximum spurious signal obtained for all background template variations.
%

%
%
%
%
%
%
%
%
%
%
%
%
%
%
%
%
%
%
%
%
%


%
\section{Statistical analysis}
\label{sec:stats}

The presence of a \signalprocess~signal is tested by means of a likelihood function built from the observed diphoton invariant mass distribution and the analytic functions discussed in Sections~\ref{sec:signal} and~\ref{sec:bkg}, describing the signal and background components in the \mgg\ range 6--68~\GeV.
The mass range is chosen to avoid bias in the background modelling due to the effects of the $\ptgg>50$~\GeV selection.
The search is then performed in the 10--60~\GeV mass range to avoid edge effects.
The parameter of interest to be extracted from the likelihood fit is the branching ratio~\br, which includes the Higgs boson production cross-section via the ggF process, $48.5 \pm 2.4$~pb~\cite{LHC-habdbook}, the dominant mechanism, and the total signal selection efficiency estimated from simulated signal samples.
The efficiency accounts for the branching ratio for hadronically decaying $\tau$-leptons, and it ranges from $2.2\times10^{-4}$ to $1.0\times10^{-4}$ for masses between 10 and 60~\GeV.
The mass dependence arises from the boosted and resolved selections --- with the former being more efficient in the low-mass regime and the latter in the high-mass regime --- as well as from the boosted topology required for the diphoton system via the \ptgg~selection.

The systematic uncertainties are implemented in the likelihood function as nuisance parameters constrained by Gaussian penalty terms, and are summarised in Table~\ref{tab:systs}.
The theoretical uncertainties affecting the measurement of \br~arise from variations of the renormalisation and factorisation scales, as well as the choice of the PDFs, modifying the total signal selection efficiency evaluated in simulated samples.
The experimental uncertainties directly impacting the signal yield include those related to the integrated luminosity, the modelling of pile-up interactions in simulation, the trigger efficiency, the photon identification and isolation efficiencies, the photon energy scale and resolution, the boosted \ditau\ object, and the resolved $\tau$-lepton reconstruction and identification efficiencies, as well as the $\tau$-lepton energy scale.
The boosted \ditau\ object uncertainty combines the \ditau\ scale factor uncertainty --- accounting for the \ditau\ reconstruction, energy scale calibration, and the statistical uncertainty of the scale factor measurement --- with the extrapolation uncertainty.
An additional systematic uncertainty in the trigger is included to account for the capability of the trigger system to identify two closely spaced electromagnetic showers, as done in Ref.~\cite{HIGG-2019-23}.
Systematic uncertainties accounting for signal and background mismodelling are included, primarily arising from the signal shape parameterisation and the fit bias (discussed in Section~\ref{sec:bkg}).
All systematic uncertainties, except those specific to either the boosted or resolved $\tau$-lepton reconstruction, are applied to both categories.

The best-fit branching ratio is obtained by performing an unbinned likelihood fit to the data under the signal-plus-background hypothesis. The compatibility of the observed data and the background-only hypothesis for a given signal hypothesis \ma\ is tested by estimating a local $p$-value based on a profile-likelihood-ratio test statistic~\cite{Cowan:2010js}.

In the absence of a signal, the expected and observed 95\% confidence level (CL) exclusion limits on the branching ratio are evaluated using the modified frequentist approach CL$_\text{s}$~\cite{Junk:1999kv} with the asymptotic approximation to the test-statistic distribution~\cite{Cowan:2010js}.
The asymptotic approximation was validated with pseudo-experiments and agrees within 10\%.

\newenvironment{DIFnomarkup}{}{}
\begin{DIFnomarkup}

\begin{table}
\centering
\caption{Summary of the main sources of systematic uncertainty.
Their impact on the branching ratio is shown, except for the background modelling uncertainty, which is expressed both as a number of events and relative to the expected statistical uncertainty $\delta S$ of a fitted signal.
The \ma-dependent uncertainties are given for two \ma\ values, 10~\GeV and 50~\GeV.
}
\small

\begin{tabular}{>{\raggedright\arraybackslash}p{8cm}
>{\centering\arraybackslash}p{2.7cm}
>{\centering\arraybackslash}p{2.7cm}}

\hline \hline
Source                                          & \multicolumn{2}{c}{Uncertainty}   \\
\hline\hline
\multicolumn{3}{c}{In \br\ [\%]}  \\
\hline
& $\ma=10~\GeV$ & $\ma=50~\GeV$ \\
\hline
Boosted \ditau\ object                          & 63 & 0.8 \\
Theory                                          & 9.9 & 27 \\
Pile-up reweighting                             & \multicolumn{2}{c}{4.5} \\
Resolved $\tau$ reconstruction, identification and energy scale     & 0.3 & 4.0 \\
Photon energy resolution                        & \multicolumn{2}{c}{3.0} \\
Photon identification efficiency                & \multicolumn{2}{c}{2.9} \\
Signal shape modelling                          & \multicolumn{2}{c}{2.5} \\
Photon isolation efficiency                     & \multicolumn{2}{c}{2.4} \\
Photon trigger efficiency                       & \multicolumn{2}{c}{$1.1$} \\
Photon energy scale                             & \multicolumn{2}{c}{$<1.0$} \\
Luminosity                                      & \multicolumn{2}{c}{0.8} \\ %
Trigger on closely spaced photons               & 0.8 & $<0.1$ \\
\hline
\multicolumn{3}{c}{In background modelling}  \\
\hline
\multirow{2}{*}{Spurious signal} & \multirow{2}{*}{\parbox{2.7cm}{\centering $<0.08 \, \sigma_{\text{stat}}$ \\ $0.16$ events}} & \multirow{2}{*}{\parbox{2.7cm}{\centering $<0.01 \, \sigma_{\text{stat}}$ \\ $0.06$ events}} \\
\\
\hline \hline
\end{tabular}
\label{tab:systs}
\end{table}

\end{DIFnomarkup}


\section{Results}
\label{sec:results}

The diphoton invariant mass distribution of events satisfying the analysis selections is shown in Figure~\ref{fig:inv_mass_data}, along with the background-only fit performed in the \mgg\ range 6--68~\GeV.

\begin{figure}
\centering
\includegraphics[width=0.7\columnwidth]{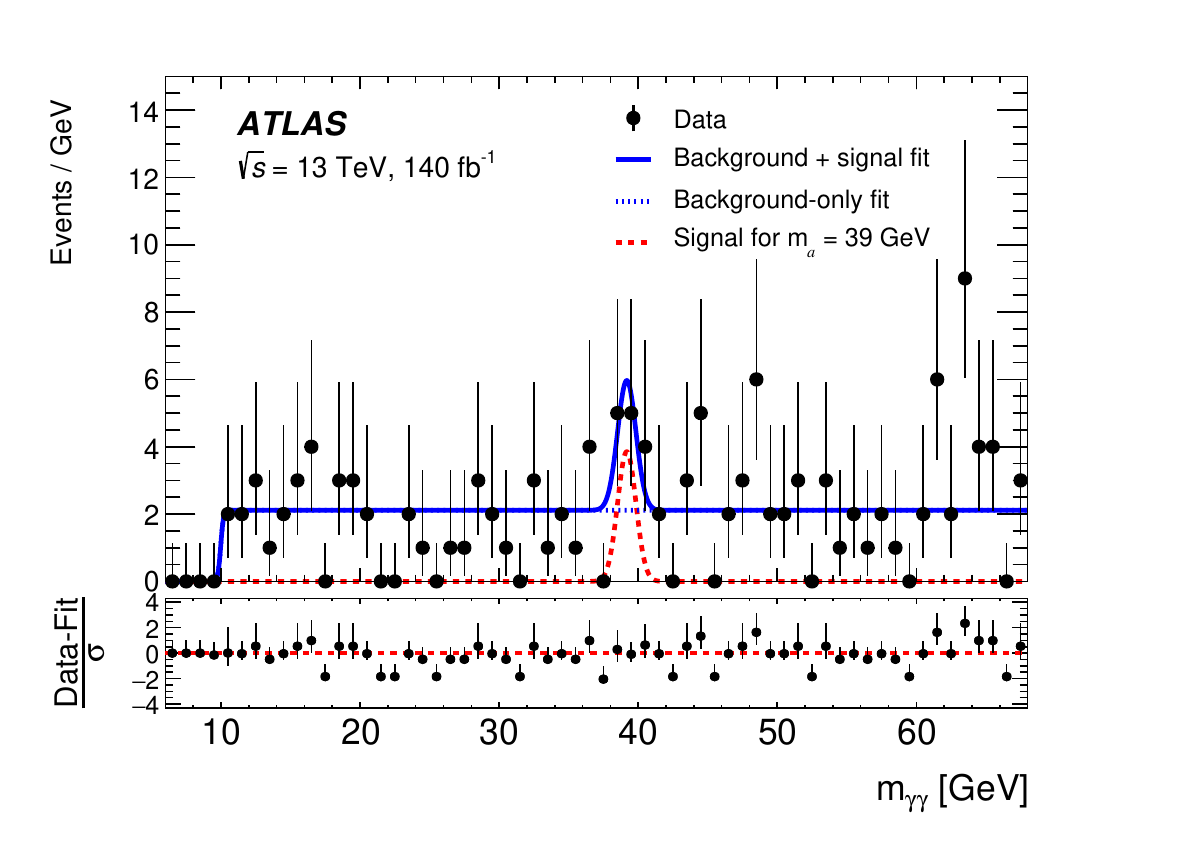}
\hfill
\caption{
Distribution of the diphoton invariant mass for all events satisfying the analysis selections in the full Run~2 dataset with the background-only fit superimposed (dotted line).
The background-plus-signal fit is given as an example for $\ma=39$~\GeV (solid line), and the corresponding signal hypothesis is shown (dashed line).
The bottom panel shows the difference between the data points and the background-plus-signal fit, expressed in units of the statistical uncertainty in the data.
}
\label{fig:inv_mass_data}
\end{figure}

The result of the $p$-value scan as a function of the hypothesised resonance mass \ma~is shown in Figure~\ref{fig:pval_scan}.
The most significant deviations from the background-only hypothesis are observed for masses of 39~\GeV and 48~\GeV, and correspond to significances of $2.2\sigma$ and $2.1\sigma$, respectively.

\begin{figure}
\centering
\subfloat[]{
\includegraphics[width=0.49\columnwidth]{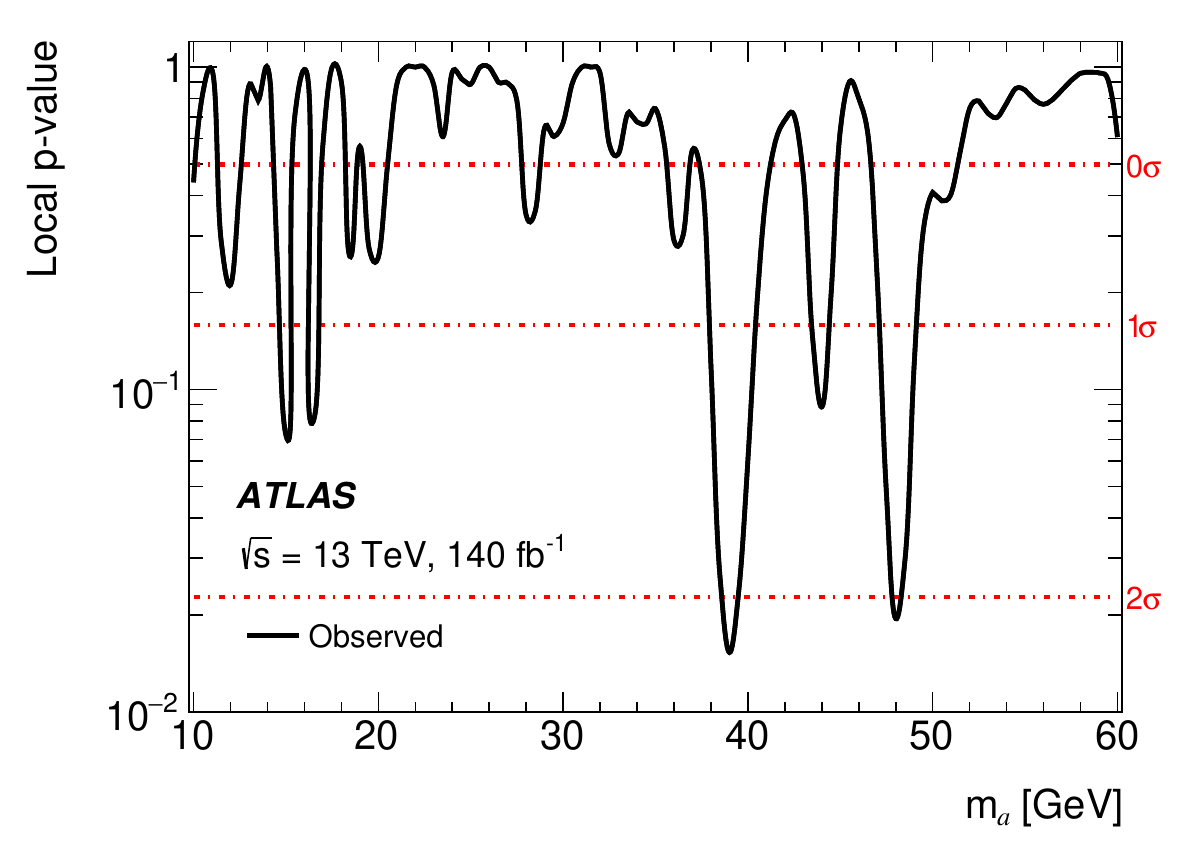}\label{fig:pval_scan}}
\hfill
\subfloat[]{
\includegraphics[width=0.49\columnwidth]{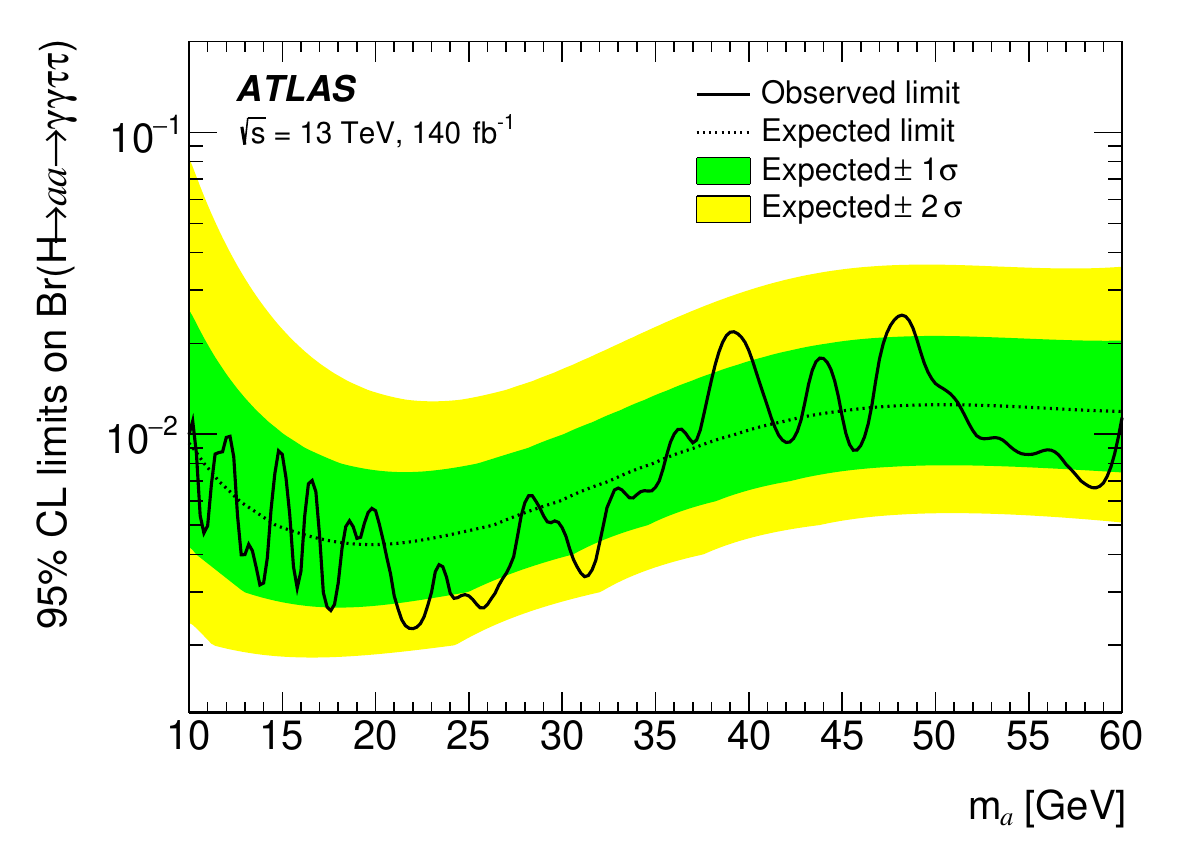}\label{fig:scan_mass}}
\hfill
\caption{
(a) Scan of the observed $p$-value as a function of \ma~for the background-only hypothesis.
(b) Observed (solid line) and expected (dashed line) upper limits at 95\% CL on the branching ratio for \signalprocessnohad~as a function of \ma. The surrounding shaded bands correspond to $\pm1$ and $\pm2$ standard deviations around the expected limit.
}
\end{figure}

The observed and expected 95\% CL upper limits on \br~as a function of \ma\ are shown in Figure~\ref{fig:scan_mass}, ranging from 0.2\% to 2\% and 0.5\% to 1\%, respectively, where the best sensitivity is achieved in \ma\ range 10--35~\GeV.
The dominant uncertainties arise from the limited number of $pp$ collisions collected, partially accounted for in the boosted \ditau\ identification uncertainty, and from the \ditau\ extrapolation uncertainty addressing the difference in the kinematic regimes probed in this analysis and in the \ditau\ calibration study.

%


%

%
%
\FloatBarrier

\section{Conclusions}
\label{sec:conclusion}

A search for SM Higgs boson decays into two pseudoscalars, $a$, with one decaying into a photon pair and the other into a hadronically decaying $\tau$-lepton pair, \signalprocess, is performed using 140~\ifb\ of $pp$ collision data at $\sqrt{s} = 13$~\TeV~collected with the ATLAS detector at the LHC\@.
This analysis considers both resolved and boosted \ditau\ topologies, using for the first time a novel algorithm for reconstructing a collimated $\tau$-lepton pair decaying hadronically, enhancing the search sensitivity at low values of the pseudoscalar mass.
The diphoton invariant mass spectrum, ranging from 10 to 60~\GeV, is analyzed to search for a narrow resonance on a smooth, quasi-flat background modelled by an analytic functional form.
The data is consistent with the SM background expectation, and limits are set on the process branching ratio.
The observed (expected) 95\% confidence level upper limits on \br~range from 0.2\% (0.5\%) to 2\% (1\%), with variations mainly due to statistical fluctuations of the data.
As the first such search conducted at the LHC, it contributes to the broad program of searches for $H \to aa$ decays that is underway, probing different final states.
%


%

%
%
%
%
%
%
\section*{Acknowledgements}
%

%

%
%

%
%

We thank CERN for the very successful operation of the LHC and its injectors, as well as the support staff at
CERN and at our institutions worldwide without whom ATLAS could not be operated efficiently.

The crucial computing support from all WLCG partners is acknowledged gratefully, in particular from CERN, the ATLAS Tier-1 facilities at TRIUMF/SFU (Canada), NDGF (Denmark, Norway, Sweden), CC-IN2P3 (France), KIT/GridKA (Germany), INFN-CNAF (Italy), NL-T1 (Netherlands), PIC (Spain), RAL (UK) and BNL (USA), the Tier-2 facilities worldwide and large non-WLCG resource providers. Major contributors of computing resources are listed in Ref.~\cite{ATL-SOFT-PUB-2025-001}.

We gratefully acknowledge the support of ANPCyT, Argentina; YerPhI, Armenia; ARC, Australia; BMWFW and FWF, Austria; ANAS, Azerbaijan; CNPq and FAPESP, Brazil; NSERC, NRC and CFI, Canada; CERN; ANID, Chile; CAS, MOST and NSFC, China; Minciencias, Colombia; MEYS CR, Czech Republic; DNRF and DNSRC, Denmark; IN2P3-CNRS and CEA-DRF/IRFU, France; SRNSFG, Georgia; BMBF, HGF and MPG, Germany; GSRI, Greece; RGC and Hong Kong SAR, China; ICHEP and Academy of Sciences and Humanities, Israel; INFN, Italy; MEXT and JSPS, Japan; CNRST, Morocco; NWO, Netherlands; RCN, Norway; MNiSW, Poland; FCT, Portugal; MNE/IFA, Romania; MSTDI, Serbia; MSSR, Slovakia; ARIS and MVZI, Slovenia; DSI/NRF, South Africa; MICIU/AEI, Spain; SRC and Wallenberg Foundation, Sweden; SERI, SNSF and Cantons of Bern and Geneva, Switzerland; NSTC, Taipei; TENMAK, T\"urkiye; STFC/UKRI, United Kingdom; DOE and NSF, United States of America.

Individual groups and members have received support from BCKDF, CANARIE, CRC and DRAC, Canada; CERN-CZ, FORTE and PRIMUS, Czech Republic; COST, ERC, ERDF, Horizon 2020, ICSC-NextGenerationEU and Marie Sk{\l}odowska-Curie Actions, European Union; Investissements d'Avenir Labex, Investissements d'Avenir Idex and ANR, France; DFG and AvH Foundation, Germany; Herakleitos, Thales and Aristeia programmes co-financed by EU-ESF and the Greek NSRF, Greece; BSF-NSF and MINERVA, Israel; NCN and NAWA, Poland; La Caixa Banking Foundation, CERCA Programme Generalitat de Catalunya and PROMETEO and GenT Programmes Generalitat Valenciana, Spain; G\"{o}ran Gustafssons Stiftelse, Sweden; The Royal Society and Leverhulme Trust, United Kingdom.

In addition, individual members wish to acknowledge support from Armenia: Yerevan Physics Institute (FAPERJ); CERN: European Organization for Nuclear Research (CERN DOCT); Chile: Agencia Nacional de Investigaci\'on y Desarrollo (FONDECYT 1230812, FONDECYT 1230987, FONDECYT 1240864); China: Chinese Ministry of Science and Technology (MOST-2023YFA1605700, MOST-2023YFA1609300), National Natural Science Foundation of China (NSFC - 12175119, NSFC 12275265, NSFC-12075060); Czech Republic: Czech Science Foundation (GACR - 24-11373S), Ministry of Education Youth and Sports (FORTE CZ.02.01.01/00/22\_008/0004632), PRIMUS Research Programme (PRIMUS/21/SCI/017); EU: H2020 European Research Council (ERC - 101002463); European Union: European Research Council (ERC - 948254, ERC 101089007, ERC, BARD, 101116429), European Union, Future Artificial Intelligence Research (FAIR-NextGenerationEU PE00000013), Italian Center for High Performance Computing, Big Data and Quantum Computing (ICSC, NextGenerationEU); France: Agence Nationale de la Recherche (ANR-20-CE31-0013, ANR-21-CE31-0013, ANR-21-CE31-0022, ANR-22-EDIR-0002); Germany: Baden-Württemberg Stiftung (BW Stiftung-Postdoc Eliteprogramme), Deutsche Forschungsgemeinschaft (DFG - 469666862, DFG - CR 312/5-2); Italy: Istituto Nazionale di Fisica Nucleare (ICSC, NextGenerationEU), Ministero dell'Università e della Ricerca (PRIN - 20223N7F8K - PNRR M4.C2.1.1); Japan: Japan Society for the Promotion of Science (JSPS KAKENHI JP22H01227, JSPS KAKENHI JP22H04944, JSPS KAKENHI JP22KK0227, JSPS KAKENHI JP23KK0245); Norway: Research Council of Norway (RCN-314472); Poland: Ministry of Science and Higher Education (IDUB AGH, POB8, D4 no 9722), Polish National Agency for Academic Exchange (PPN/PPO/2020/1/00002/U/00001), Polish National Science Centre (NCN 2021/42/E/ST2/00350, NCN OPUS 2023/51/B/ST2/02507, NCN OPUS nr 2022/47/B/ST2/03059, NCN UMO-2019/34/E/ST2/00393, NCN \& H2020 MSCA 945339, UMO-2020/37/B/ST2/01043, UMO-2021/40/C/ST2/00187, UMO-2022/47/O/ST2/00148, UMO-2023/49/B/ST2/04085, UMO-2023/51/B/ST2/00920); Spain: Generalitat Valenciana (Artemisa, FEDER, IDIFEDER/2018/048), Ministry of Science and Innovation (MCIN \& NextGenEU PCI2022-135018-2, MICIN \& FEDER PID2021-125273NB, RYC2019-028510-I, RYC2020-030254-I, RYC2021-031273-I, RYC2022-038164-I); Sweden: Carl Trygger Foundation (Carl Trygger Foundation CTS 22:2312), Swedish Research Council (Swedish Research Council 2023-04654, VR 2018-00482, VR 2022-03845, VR 2022-04683, VR 2023-03403, VR grant 2021-03651), Knut and Alice Wallenberg Foundation (KAW 2018.0458, KAW 2019.0447, KAW 2022.0358); Switzerland: Swiss National Science Foundation (SNSF - PCEFP2\_194658); United Kingdom: Leverhulme Trust (Leverhulme Trust RPG-2020-004), Royal Society (NIF-R1-231091); United States of America: U.S. Department of Energy (ECA DE-AC02-76SF00515), Neubauer Family Foundation.

%
%


%
%
%
%
%
%

%
%
%
\printbibliography
\clearpage
 
\begin{flushleft}
\hypersetup{urlcolor=black}
{\Large The ATLAS Collaboration}

\bigskip

\AtlasOrcid[0000-0002-6665-4934]{G.~Aad}$^\textrm{\scriptsize 104}$,
\AtlasOrcid[0000-0001-7616-1554]{E.~Aakvaag}$^\textrm{\scriptsize 17}$,
\AtlasOrcid[0000-0002-5888-2734]{B.~Abbott}$^\textrm{\scriptsize 123}$,
\AtlasOrcid[0000-0002-0287-5869]{S.~Abdelhameed}$^\textrm{\scriptsize 119a}$,
\AtlasOrcid[0000-0002-1002-1652]{K.~Abeling}$^\textrm{\scriptsize 56}$,
\AtlasOrcid[0000-0001-5763-2760]{N.J.~Abicht}$^\textrm{\scriptsize 50}$,
\AtlasOrcid[0000-0002-8496-9294]{S.H.~Abidi}$^\textrm{\scriptsize 30}$,
\AtlasOrcid[0009-0003-6578-220X]{M.~Aboelela}$^\textrm{\scriptsize 46}$,
\AtlasOrcid[0000-0002-9987-2292]{A.~Aboulhorma}$^\textrm{\scriptsize 36e}$,
\AtlasOrcid[0000-0001-5329-6640]{H.~Abramowicz}$^\textrm{\scriptsize 155}$
% ...
\bigskip
\\

% ...
$^{17}$Department for Physics and Technology, University of Bergen, Bergen; Norway.\\
% ...
$^{30}$Physics Department, Brookhaven National Laboratory, Upton NY; United States of America.\\
% ...
$^{36}$$^{(a)}$Facult\'e des Sciences Ain Chock, Universit\'e Hassan II de Casablanca;$^{(b)}$Facult\'{e} des Sciences, Universit\'{e} Ibn-Tofail, K\'{e}nitra;$^{(c)}$Facult\'e des Sciences Semlalia, Universit\'e Cadi Ayyad, LPHEA-Marrakech;$^{(d)}$LPMR, Facult\'e des Sciences, Universit\'e Mohamed Premier, Oujda;$^{(e)}$Facult\'e des sciences, Universit\'e Mohammed V, Rabat;$^{(f)}$Institute of Applied Physics, Mohammed VI Polytechnic University, Ben Guerir; Morocco.\\
% ...
$^{46}$Physics Department, Southern Methodist University, Dallas TX; United States of America.\\
% ...
$^{50}$Fakult\"{a}t Physik , Technische Universit{\"a}t Dortmund, Dortmund; Germany.\\
% ...
$^{56}$II. Physikalisches Institut, Georg-August-Universit\"{a}t G\"ottingen, G\"ottingen; Germany.\\
% ...
$^{104}$CPPM, Aix-Marseille Universit\'e, CNRS/IN2P3, Marseille; France.\\
% ...
$^{119}$$^{(a)}$New York University Abu Dhabi, Abu Dhabi;$^{(b)}$United Arab Emirates University, Al Ain; United Arab Emirates.\\
% ...
$^{123}$Homer L. Dodge Department of Physics and Astronomy, University of Oklahoma, Norman OK; United States of America.\\
% ...
$^{155}$Raymond and Beverly Sackler School of Physics and Astronomy, Tel Aviv University, Tel Aviv; Israel.\\
% ...
$^{a}$ Also Affiliated with an institute covered by a cooperation agreement with CERN.\\
$^{b}$ Also at An-Najah National University, Nablus; Palestine.\\
$^{c}$ Also at Borough of Manhattan Community College, City University of New York, New York NY; United States of America.\\
$^{d}$ Also at Center for High Energy Physics, Peking University; China.\\
$^{e}$ Also at Center for Interdisciplinary Research and Innovation (CIRI-AUTH), Thessaloniki; Greece.\\
$^{f}$ Also at CERN, Geneva; Switzerland.\\
% ...

\end{flushleft}


%

%
%

%

%
%
%
%

\end{document}